\documentclass[showpacs,preprintnumbers,amsmath,amssymb]{revtex4}
\usepackage{dcolumn}
\usepackage{bm}
\usepackage{epsfig}
\usepackage{color}

\newcommand{\pa}{\partial}

\renewcommand{\bar}[1]{\overline{#1}}
\newcommand{\ie}{{\it i.e.}}

\usepackage{amssymb}

\renewcommand{\bar}[1]{\overline{#1}}

\usepackage{indentfirst}
\usepackage{psfig,color}
\usepackage{epsfig}
\usepackage{graphicx}

\providecommand{\Journal}[4] {#1 {\bf #2}, #3 (#4)}
\providecommand{\NPA}{Nucl. Phys. A } %
\providecommand{\NPB}{Nucl. Phys. B } %
\providecommand{\PLB}{Phys. Lett. B } %
\providecommand{\PRL}{Phys. Rev. Lett. } %
\providecommand{\PRC}{Phys. Rev. C } %
\providecommand{\PRD}{Phys. Rev. D } %
\providecommand{\PRSA}{Proc. Roy. Soc. Lond. A } %
\providecommand{\RMP}{Rev. Mod. Phys. } %
\providecommand{\RMP}{Rev. Mod. Phys. } %

\begin{document}

\title{Baryonium with a phenomenological skyrmion-type potential}

\author{Mu-Lin Yan}
\author{Si Li}
\affiliation{Interdisciplinary Center for Theoretical Study,
University of Science and Technology of China, Hefei, Anhui
230026, China}
\author{Bin Wu}
\author{Bo-Qiang Ma}
\email{mabq@phy.pku.edu.cn}\altaffiliation{corresponding author.}
\affiliation{Department of Physics, Peking University, Beijing
1000871, China}


\begin{abstract}
In this paper, we investigate the nucleon-antinucleon static
energies in the Skyrme model with the product {\it Anzatz}. The
calculation shows that in the ungroomed $S\bar S$ (skyrmion and
antiskyrmion) channel, which leads to rapid annihilation, there
exists a quasi-stable bound state which may give a natural
explanation for the near-threshold enhancement in the
proton-antiproton $(p\bar{p})$ mass spectrum reported by the BES
Collaboration and the Belle Collaboration. Similar to the
phenomenological well potential of the deuteron, we construct a
phenomenological skyrmion-type potential to study this narrow
$p\bar{p}$-resonance in $ J/\psi\rightarrow \gamma p\bar{p}$. By
this potential model, a $p\bar{p}$ baryonium with small binding
energies is suggested and the decay width of this state is
calculated by WKB approximation. In this picture the decay is
attributed to quantum tunnelling and $p\bar{p}$ annihilation.
Prediction on the decay mode from the baryonium annihilation at
rest is also pointed out.
\end{abstract}

\pacs{12.39.Pn, 13.75.Cs, 12.39.Dc, 12.39.Mk}

\maketitle

\section{Introduction}

Skyrme's old idea~\cite{Skyr} that baryons are chiral solitons has
been successful in describing the static nucleon
properties~\cite{Adkins} since Witten's illustration that the
soliton picture of baryons is consistent with QCD in the large
$N_c$ approximation~\cite{witt}. The Skyrme model has been widely
used to discuss baryons and baryonic-system properties. The Skyrme
model with the product {\it Ansatz} has also been applied to the
nucleon-nucleon interaction~\cite{jjp} and to the static
properties of the deuteron~\cite{b_c}. At the classical level, the
deuteron mass has been successfully calculated in {\it ab initio}
approach~\cite{Schramm}. Atiyah and Manton~\cite{a_m} revealed a
mathematically elegant connection between the Skyrme soliton and
the instantons in the SU(2) gauge field, in which the holonomy of
the instanton along all lines parallel to the Euclidean time axis
generates the skyrmion with the same topological charge. Using the
{\it Anzatz} generated in this approach, Leese {\it
et~al.}~\cite{lms} reinvestigated the deuteron properties and gave
an experimentally more agreeable result. However, since self-dual
and antiself-dual equations allow only trivial solutions of zero
topological charge, it is difficult to apply the Skyrme model to
the interaction between skyrmion and anti-skyrmion in such a
manner. The nucleon-antinucleon static potential was studied in
the Skyrme model using the product {\it Anzatz} by Lu and
Amado~\cite{lu_am}, who showed that the Skyrme picture with the
product {\it Ansatz} is the reasonable first step to obtain the
real part of nucleon-antinucleon interaction.

Recently, the BES Collaboration observed a near-threshold
enhancement in the proton-antiproton $(p\bar{p})$ mass spectrum
from the radiative decay $J/\psi\rightarrow \gamma p\bar{p}$
\cite{BES}. This enhancement can be fitted with either an $S$- or
$P$-wave Breit-Wigner resonance function. In the case of $S$-wave
fit, the peak mass is at $M=1859_{-10}^{+3}({\rm
stat})_{-25}^{+5}({\rm sys})$ with a total width $\Gamma<
30~\mbox{MeV/c}^2$ at $90\%$ percent confidence level. The
corresponding spin and parity are $J^{PC}=0^{-+}$.  Moreover, the
Belle Collaboration also reported similar observations of the
decays $B^+\rightarrow K^+p\bar{p}$ \cite{Abe1} and
$\bar{B}^0\rightarrow D^0p\bar{p}$ \cite{Abe2}, showing
enhancements in the $p\bar{p}$ invariant mass distributions near
$2m_p$. These observations could be interpreted as signals for
baryonium $p\bar{p}$ bound states~\cite{Datta} or flavorless gluon
states~\cite{Rosner}. There are also suggestions that the
near-threshold enhancement of $p\bar{p}$ is due to final state
interactions~\cite{Zou} or as a result of the quark fragmentation
process~\cite{Rosner}.

In this paper, we provide a possible explanation that the
enhancement might be explained as a baryonium $p\bar{p}$ bound
state in a phenomenological potential inspired by the
investigation of the static energy in the ungroomed $S\bar S$
channel. In Sec.~2, we scrutinize the nucleon-antinucleon static
energy in the ungroomed $S\bar S$ channel from the $SU(2)$ Skyrme
model. In Sec.~3, inspired by this energy, we construct a
phenomenological skyrmion-type potential, investigate the bound
state in this potential and calculate the width by WKB
approximation through quantum tunnelling effect. Then, in Sec.~4,
we give our conclusion and discussion, with emphasis on the
significant implication on the decay mode of the baryonium
annihilation at rest.

\section{The static energy in the ungroomed $S\bar S$ channel}
The Lagrangian for the SU(2) Skyrme model is
\begin{equation}
    {\mathcal{L}}
    =\frac{1}{16}F^2_\pi {\mbox{Tr}}(\partial_\mu U^{\dagger} \partial^\mu
    U)+\frac{1}{32e^2}{\mbox{Tr}}([(\partial_\mu U) U^{\dagger},(\partial_\nu
    U) U^{\dagger}]^2)+\frac{1}{8}m^2_\pi F^2_\pi
    {\mbox{Tr}}(U-1),~~
\end{equation}
 where  $U(t,\textbf{x})$ is the SU(2) chiral field, expressed in
 terms of the pion fields:
 \begin{equation}
    U(t,\mathbf{x})=\sigma(t,\mathbf{x})+i\pi(t,\mathbf{x})\cdot\mathbf{\tau}.
 \end{equation}
We fix the parameters $F_\pi$ and $e$ as in \cite{Adkins}, and our
units are related to conventional units via
\begin{equation}
\frac{F_{\pi}}{4e}=5.58~\mbox{MeV},     \ \
\frac{2}{eF_{\pi}}=0.755~\mbox{fm}.
\end{equation}
Skyrme avoided the obstacle of minimizing the energy by invoking
the hedgehog {\it Ansatz}
\begin{equation}
        U_{H}(\mathbf{r})=\exp[i\tau\cdot\widehat{r}f(r)].
\end{equation}
The product {\it Ansatz} describing the behavior of skyrmion and
antiskyrmion system with relatively arbitrary rotation in the
iso-space is of the form
\begin{equation}
    U_s=CU_{H}(\mathbf{r}+\frac{\rho}{2}\hat{z})C^\dagger
    U_{H}^\dagger(\mathbf{r}-\frac{\rho}{2}\hat{z}),
\end{equation}
where $C$ is an element in the isospin SU(2) group. The skyrmion
and the antiskyrmion are separated along the $\hat{z}$-axis by a
distance $\rho$. The interactions between various  groomed
skyrmion and antiskyrmion have been studied in Ref.~\cite{lu_am},
in which the configuration lies in a manifold of relatively higher
dimension. We find that the static potential in the ungroomed
$S\bar S$ channel is physically more interesting with $C=I$, which
satisfies
\begin{equation}
    U\rightarrow 1 \ \ \mbox{when} \ \ \rho\rightarrow 0.
\end{equation}
Fixing our {\it Ansatz} as above, we can get the static energy
from the skyrmion Lagrangian:
\begin{equation}\label{V}
    M(\rho)=\int d^3{\bf r}[
    -\frac{1}{2}{\mbox{Tr}}(R_{i}R_{i})-\frac{1}{16}{\mbox{Tr}}([R_{i},R_{j}]^2)-m_\pi^2\mbox{Tr}(U-1)],
\end{equation}
where $i,j=1,2,3$. The right-currents $R_{\mu}$ are defined via
\begin{equation}
    R_{\mu}=(\partial_{\mu}U)U^\dagger,
\end{equation}
and we express the energy in the units defined above. In this
picture, the binding energies for $S\bar{S}$ (which correspond to
the classical binding energies of $p\bar{p}$) are
\begin{equation}\label{binding}
\Delta E_B=2m_p^c-M(\rho),
\end{equation}
where $m_p^c=867~\mbox{MeV}$ is the mass of a classical nucleon
(or classical skyrmion). A stable or quasi-stable
$p\bar{p}$-binding state corresponds to the skyrmion configuration
$U_s(r,\rho_B)=U_{H}(\mathbf{r}+\frac{\rho_B}{2}\hat{z})
    U_{H}^\dagger(\mathbf{r}-\frac{\rho_B}{2}\hat{z})$ with
$\Delta  E_B(\rho_B)<0$ and ${d \over d\rho}(\Delta
E_B(\rho))|_{\rho=\rho_B}=0.$

The numerical result of the static energy as a function of $\rho$
is showed in Fig.~1. From it, we find that there is a quasi-stable
$p\bar{p}$-binding state:
\begin{eqnarray}\label{B1}
\rho_B &\approx& 2.5~\mbox{fm}, \\
\label{B2} \Delta E_B(\rho_B) &\approx& 10~\mbox{MeV}.
\end{eqnarray}

In the above investigation, we actually conjecture the binding
energy between $S$ and $\overline{S}$ under the {\it Anzatz} of
$U_s(r,\rho_B)=U_{H}(\mathbf{r}+\frac{\rho_B}{2}\hat{z})
    U_{H}^\dagger(\mathbf{r}-\frac{\rho_B}{2}\hat{z})$ being approximately the
 binding energy of $p\bar{p}$. This conjecture is based on the
considerations in the studies of deuteron as a soliton in the
Skyrme model done by Braaten and Carson~\cite{b_c}. In their
works, the two-skyrmion configuration with $B=2$ has been treated
as a single soliton $U_2({\bf r})$ by using the product {\it
Ansatz}, and the spin- and isospin-quantum numbers of the physical
states arise from the semiclassical quantization of the collective
coordinates of the soliton. By this way, the deuteron state with
($I=0,~J=1$) has been identified, and its mass (or its
corresponding binding energy) has been obtained. In this
formulation, the deuteron's mass (or energies) is divided into two
parts: (1) the classical energies of statical soliton
configuration $U_2({\bf r})$; (2) the semiclassical corrections.
The latter one is suppressed in large $N_c$-expansion limit.
Namely, the classical part of the energies of the skyrmion, \ie,
the toroidal configuration is qualitatively dominant. Moreover,
the detailed analysis by Forest et al. does indeed exhibit the
toroidal configuration in their quantum mechanical deuteron wave
function, which, though making up only a very small component,
could have a revealing relation to certain aspects of
QCD~\cite{Pand}. We argue that when one uses similar product {\it
Ansatz} to deal with the skyrmion-antiskyrmion system, the
situation should be similar to the case of skyrmion-skyrmion .
Namely, the $(S\bar{S})$-binding energy eq.~(\ref{B2}) should
roughly be the $(p\bar{p})$-binding energy. And the spin- and
isospin-quantum numbers of the $(p\bar{p})$-bound state arise from
the semiclassical quantization of the collective coordinates of
$U_s(r, \rho_B)$. Since $U_s(r, \rho_B)$ serves as a classical
soliton {\it Ansatz} of $(p\bar{p})$ (rather than $(NN)$ for
deuteron case), it could be expected that the state with spin-0
and isospin-0 would be the lowest semi-classical state. However,
the real situation might be complicated and we leave this subject
to be one of the topics for future studies.

Of course, we should not take this potential too seriously for two
reasons. On one hand, the product {\it Anzatz} is an exact
solution to the equation of motion of the system only when $\rho_B
\rightarrow\infty$. On the other hand, even in the case of
Deuteron~\cite{b_c,lms}, the toroidal configuration gives a
binding energy of the order of the strong interaction scale, which
is $\sim$ pion decay constant ($\sim$ 100 MeV), not of the nuclear
scale, $\sim$ a few MeV, and the size and quadrupole moment are
simply too small. However, the potential in Fig.~1 seems
suggestive to give the observed near-threshold $p \bar{p}$
enhancement in experiments~\cite{BES,Abe1,Abe2} a possible
phenomenological explanation, as simple as possible, just like
that in Ref.~\cite{Datta} based on the inspiration from it.
Moreover, by such a phenomenological potential reflecting the
character of Fig.~1, we can also obtain a picture about how the
$p\bar p$ bound state decays: there is a $p\bar p$ bound state in
a well, which will mostly annihilate through a barrier penetration
by tunnel effect, and this will be discussed in detailed in the
next section.



\begin{figure}[hptb]
   \centerline{\psfig{figure=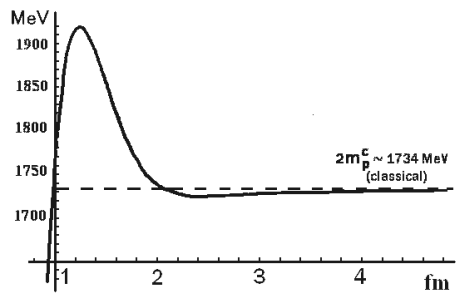,width=8cm}}
 \centering
 \vskip-0.5in
\begin{minipage}{3.45in}
    \caption{The static energy of skyrmion-antiskyrmion
    system, where $m_p^c$ is the classical single skyrmion mass without quantum correction.}
\end{minipage}
\end{figure}

\section{A phenomenological model with a skyrmion-type
potential}

In this section, we employ the phenomenological model induced from
the skyrmion picture of $(p\bar{p})$-interactions for the
nucleons. We favor such a potential as showed in Fig.~1 because it
seems that the potential can be physically substantiated. For over
fifty years there has been a general understanding of the
nucleon-nucleon interaction as one in which there is, in potential
model terms, a strong repulsive short distance core together with
a longer range weaker attraction. The attractive potential at the
middle range
 binds the neutron and the proton to form a deuteron.
In comparison with the skyrmion result on the
deuteron~\cite{b_c,a_m,lms}, we notice several remarkable features
of the static energy $M( \rho)$ and the corresponding
$(p\bar{p})$-potential $V(\rho)$ . Firstly, the potential is
attractive at $\rho>2.0~\mbox{fm}$, similar to the deuteron case.
This is due to the reason that the interaction via pseudoscalar
$\pi$-meson exchange is attractive for both quark-quark $qq$ and
quark-antiquark $q\bar{q}$ pairs. Physically, the attractive force
between $p$ and $\bar{p}$ should be stronger than that of $p n$.
Therefore the fact that our result of the $p\bar{p}$-binding
energy (see eq.(\ref{B2})) $\Delta E_B(\rho_B) (\approx
10~\mbox{MeV})$ is larger than that of deuteron
($2.225~\mbox{MeV}$) is quite reasonable physically. Secondly,
there is a static skyrmion energy peak at $\rho\sim 1~\mbox{fm}$
in Fig.~1. This means that the corresponding potential between $p$
and $\bar{p}$ is repulsive at that range. This is an unusual and
also an essential feature. The possible explanation for it is that
the skyrmions are extended objects, and there would emerge a
repulsive force to counteract the deformations of their
configuration shapes when they close to each other.  Similar
repulsive potential has also been found in previous numerical
calculation~\cite{lu_am}. Thirdly, a well potential at middle
$\rho$-range is formed due to the competition between the
repulsive and attractive potentials mentioned above, similar to
the deuteron case. But the depth should be deeper than that in the
deuteron case, as argued in a QCD based discussion~\cite{Datta}.
The $p$ and $\bar{p}$ will be bound to form a baryonium in this
well potential. Finally, the potential turns to decrease quickly
from $\sim 2000~\mbox{MeV}$ to zero when $\rho \to 0$. This means
that there is a strong attractive force at $\rho \sim 0$.
Physically, $p\bar{p}$ are annihilated.

\begin{figure}[hptb]
   \centerline{\psfig{figure=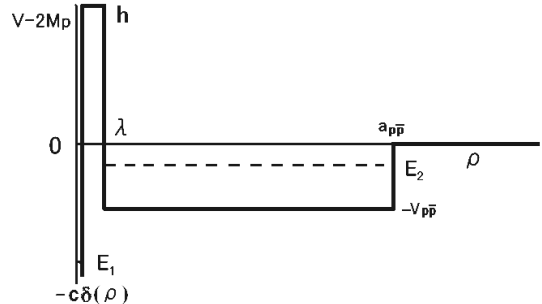,width=7.5cm}}
 \centering
\vskip-0.3in
\begin{minipage}{3.45in}
    \caption{The skyrmion-type potential of $p\bar{p}$-system.}
\end{minipage}
\end{figure}

The qualitative features of the proton-neutron potential for the
deuteron can be well described by a simple phenomenological model
of a square well potential 
\cite{Ma,Hulthen,Schiff} with a depth which is sufficient to bind
the $pn$ ${^3S_1}$-state with a binding energy of
$-2.225~\mbox{MeV}$. Numerically, the potential width $a_{pn}$ is
about $2.0~\mbox{fm}$, and the depth is about
$V_{pn}=36.5~\mbox{MeV}$. Similarly, from the above illustration
on the features of the potential between  $p$ and $\bar{p}$ based
on the Skyrmion picture, we now construct a phenomenological
potential model for the $p\bar{p}$ system, as shown in Fig.~2, and
it will be called as the skyrmion-type potential hereafter.

We take the width of the square well potential, denoted as
$a_{p\bar{p}}$, as close to that of the deuteron,  i.e.,
$a_{p\bar{p}}\sim a_{pn}\simeq 2.0 ~\mbox{fm}$. According to QCD
inspired considerations~\cite{Datta,Maltman,Rujula}, the well
potential between $q$ and $\bar{q}$ should be double attractive
than the $qq$-case, i.e., the depth of the $p\bar{p}$ square well
potential is $V_{p\overline{p}}\simeq 2V_{pn}=73~\mbox{MeV}$. The
width for the repulsive force revealed by the Skyrme model can be
fitted by the decay width of the baryonium, and we take it to be
$\lambda = 1/(2m_p) \sim 0.1~\mbox{fm}$, the Compton wave length
of the bound state of $p\bar{p}$. The square barrier potential
begins from $\rho \sim \lambda$, and the height of the potential
barrier, which should be constrained by both the decay width and
the binding energy of the baryonium, is taken as $2m_p+h$, where
$h \sim m_p/4$. At $\rho\sim 0$, $V^{(p\bar{p})}(\rho)\sim
-c\;\delta (\rho)$ with a constant $c>0$.

Analytically, the potential $V(\rho)$ is expressed as follows
\begin{equation}\label{potential}
V(\rho)=2m_p-c\;\delta(\rho)+V_c(\rho),
\end{equation}
where
\begin{equation}
V_c(\rho)=\left\{
   \begin{array}{ll}
   {h={m_p/4}},\ \ & 0<\rho<\lambda,\\
   -V_{p\bar{p}}
   =-73~\mbox{MeV},     & \lambda<\rho<a_{p\bar{p}},\\
   0, &~~~~~\rho>a_{p\bar{p}}.
   \end{array} \right.
\end{equation}
With this potential, the Schr$\ddot{o}$dinger equation for
$S$-wave bound states is
\begin{equation}\label{schrodinger}
{-1\over 2(m_p/2)}{\pa^2\over \pa \rho^2}
u(\rho)+\left[V(\rho)-E\right]u(\rho)=0,
\end{equation}
where $u(\rho)=\rho\;\psi(\rho)$ is the radial wave function,
$m_p/2$ is the reduced mass. This equation can be solved
analytically, and there are two bound states $u_1(\rho)$ and
$u_2(\rho)$: $u_1(\rho)$ with binding energy $E_1<
-V_{p\bar{p}}=-73~\mbox{MeV}$ is due to
$-c\;\delta(\rho)$-function potential mainly, and $u_2(\rho)$ with
binding energy $E_2>-73~\mbox{MeV}$ is due to the attractive
square well potential at middle range mainly.
$u_1(\rho)$ is the vacuum state, and, clearly, $u_2(\rho)$ should
correspond to a deuteron-like molecule state and it may be
interpreted as the new $p\bar{p}$ resonance reported by
BES~\cite{BES}. It is also expected that corresponding binding
energies $-E_2$ in the potential model provided in above $\Delta
E_B(\rho_B)$ (see Eq.~(\ref{B2})) are all in agreement of the data
within errors of BES~\cite{BES}. By fitting experimental data, we
have
\begin{equation}\label{E1} E_1=-(2m_p-m_{\eta_0})\simeq
-976~\mbox{MeV},
\end{equation}
\begin{equation}
E_2=-17.2~\mbox{MeV}.
\end{equation}
Considering its decay width which will be derived soon (see
Eq.~(\ref{GG})), we conclude that the near-threshold narrow
enhancement in the $p\bar{p}$ invariant mass spectrum from
$J/\psi\rightarrow\gamma p
\bar{p}$ 
might be interpreted as a state of protonium in this potential
model.

In the skyrmion-type potential of $p\bar{p}$, there are two
attractive potential wells: one is at $\rho\sim 0$ and the other
is at middle scale, together with a potential barrier between
them. At $\rho\sim 0$, the baryon- and anti-baryon pair
annihilates. The baryon-antibaryon annihilation has been studied
in the Skyrme model in literature, see, e.g.,
Refs.~\cite{somm,Amado1}. Naturally, we postulate that the bound
states decay dominantly by annihilation and, therefore, we can
derive the width of protonium state $u_2(\rho)$ by calculating the
quantum tunnelling effect for $u_2(\rho)$ passing through the
potential barrier. By WKB-approximation, the tunnelling
coefficient (i.e., barrier penetrability) reads \cite{Schiff}
\begin{eqnarray}\label{WKB}
T_0&=&\exp\left[-2\int_0^{\lambda} dr \sqrt{m_p(h-E_2)}\right]
\nonumber \\  &=&\exp \left[-2\lambda\sqrt{m_p(h-E_2)}\right].
\end{eqnarray}
In the square well potential from $\lambda$ to $a_{p\bar{p}}$, the
time-period $\theta$ of round trip for the particle is
\begin{equation}
\label{trip} \theta={2\left[a_{p\bar{p}}-\lambda\right]\over
v}=\left[a_{p\bar{p}}-\lambda\right]\sqrt{{m_p\over
V_{p\bar{p}}+E_2}}.~
\end{equation}
Thus, the state $u_2(r)$'s life-span is $\tau=\theta T_0^{-1}$,
and hence the width of that state reads
\begin{eqnarray}
\label{Gamma} \Gamma\equiv {1\over \tau}
&=& {1\over a_{p\bar{p}}-\lambda} \sqrt{{ V_{p\bar{p}}+E_2 \over
m_p}}
\nonumber \\
&~ & \exp \left[-2\lambda\sqrt{m_p(h-E_2)}\right].~~
\end{eqnarray}
Numerically, substituting
$E_2=-17.2~\mbox{MeV},\;a_{p\bar{p}}=2.0~\mbox{fm}$ into
(\ref{Gamma}), we obtain the prediction of $\Gamma$:
\begin{equation}\label{GG}
\Gamma\simeq 15.5~\mbox{MeV},
\end{equation}
which is compatible with the experimental data \cite{BES}.


\section{Conclusion}
In conclusion, we investigated the $S\bar S$ static potential in
the Skyrme model, and similar to the phenomenological potential
model of the deuteron, we constructed a skyrmion-type potential
model to study the recent discovery of a narrow
$N\bar{N}$-resonance in the decay $J/\psi\rightarrow \gamma
p\bar{p}$ by BES, and also in the decays $B^+\rightarrow
K^+p\bar{p}$ and $\bar{B}^0\rightarrow D^0p\bar{p}$ by Belle. The
parameters in the model are guided by the parameters in the
deuteron model and by QCD inspired considerations. We found that
this  skyrmion-type potential model has one baryonium solution,
which might be explained as the $p\bar p$ bound state. By fitting
the mass, we found that the width can be compatible with the
experimental data.

We have learned from the studies in this paper that based on the
skyrmion considerations, the baryonium decays are mainly due to
the annihilations of nucleon-antinucleon. This is a significant
feature for the particle decay of the baryonium. It has been well
known that the nucleon-antinucleon annihilation at rest mostly
favors processes with 4 to 7 pions in the final states over those
with two or three pions~\cite{Amado1}. Considering that the
binding energy of the nucleon (or antinuleon) is rather small
(compared with the mass of nucleon), the annihilation of the
baryonium occurs nearly at rest. Therefore, we would predict that
the baryonium hadronic decay should also mostly favor processes
with 4 to 7 pseudoscalar mesons in the final states over those
with two or three mesons, even though the phase spaces for the
latter are larger than the former one. This prediction is
non-trivial and needs to be tested by experiments.


{\bf Acknowledgements} We acknowledge Shan Jin and Zhi-Peng Zheng
for stimulating discussions. This work is partially supported by
National Natural Science Foundation of China under Grant Numbers
90103002, 90103007, 10025523, and 10421003,  by the PhD Program
Funds of the Education Ministry of China and  KJCX2-SW-N10 of the
Chinese Academy, and by the Key Grant Project of Chinese Ministry
of Education (No.~305001).




\begin{thebibliography}{99}

\bibitem{Skyr}
T.H.R.~Skyrme, \Journal{\PRSA}{260}{127}{1961}.

\bibitem{Adkins}
G.S.~Adkins, C.R.~Nappi, E.~Witten,
\Journal{\NPB}{228}{552}{1983};
 G.S.~Adkins, C.R.~Nappi,
\Journal{\NPB}{233}{109}{1984}

\bibitem{witt}
E.~Witten, \Journal{\NPB} {223}{422 and 433}{1983}.

\bibitem{jjp}
A.~Jackson, A.D.~Jackson, V.~Pasquier,
\Journal{\NPA}{432}{567}{1985}.

\bibitem{b_c}
E.~Braaten, L.~Carson, \Journal{\PRL}{56}{1897}{1986};
\Journal{\PRD}{38}{3525}{1988}.

\bibitem{Schramm} A.J. Schramm, \Journal{\PRC}{37}{1799}{1988}.
\bibitem{a_m}
M.F.~Atiyah, N.S.~Manton, \Journal{\PLB}{222}{438}{1989};
\Journal{Commun.~Math.~Phys.}{153}{391}{1993}.

\bibitem{lms}
R.A.~Leese, N.S.~Manton, B.J.~Schroers,
\Journal{\NPB}{442}{228}{1995}.

\bibitem{Pand}
J.L.~Forest, et al., \Journal{\PRC}{54}{646}{1996}.

\bibitem{lu_am}
Y.~Lu, R.D.~Amado, \Journal{\PRC}{54}{1566}{1996}.

\bibitem{BES} J.Z.~Bai, et al., \Journal{\PRL}{91}{022001}{2003}.
\bibitem{Abe1} K.~Abe, et al., \Journal{\PRL}{88}{181803}{2002}.
\bibitem{Abe2} K.~Abe, et al., \Journal{\PRL}{89}{151802}{2002}.

\bibitem{Datta} A.~Datta, P.J.~O'Donnell, \Journal{\PLB}{567}{273}{2003}.

\bibitem{Rosner}
J.L.~Rosner, \Journal{\PRD}{68}{014004}{2003}.

\bibitem{Zou} B.S.~Zou,
H.C.~Chiang, \Journal{\PRD}{69}{034004}{2004}; B.~Kerbikov,
A.~Stavinsky, V.~Fedotov, \Journal{\PRC}{69}{055205}{2004}.



\bibitem{Ma}
S.~Ma, \Journal{\RMP}{25}{853}{1953}.
\bibitem{Hulthen}
L.~Hulthen, M.~Sugawara, \Journal{Handbuch der
Physik}{39}{1}{1959}.
\bibitem{Schiff}
L.I.~Schiff, {\it Quantum Mechanics} (McGraw-Hill Book Company,
Inc., 1949).

\bibitem{Maltman} K.~Maltman, N.~Isgur, \Journal{\PRD}{29}{952}{1984}.


\bibitem{Rujula} A.~De Rujula, H.~Georgi, S.L.~Glashow, \Journal{\PRD}{12}{147}{1975}.
\bibitem{somm} H.M.~Sommermann, R.~Seki, S.~Larson and S.E.~Koonin,
 \Journal{\PRD}{45}{4303}{1992}.
\bibitem{Amado1} R.D. Amado, Nucl. Phys.
B, Proc. Suppl. {\bf A56}, 22 (1997); Nucl. Phys. {\bf A655}, 236c
(1999).






\end{thebibliography}
\end{document}